\documentclass[aps,pra,twocolumn,showpacs,floatfix]{revtex4}
\usepackage{graphicx}

\begin{document}


\title{Stability and entanglement in optical-atomic amplification of trapped
atoms: the role of atomic collisions}

\author{V. V. Fran\c{c}a$^{1,3}$ and G. A. Prataviera$^{2,3,}$}
\email{gap@df.ufscar.br}
\affiliation{\\
$^{1}$Departamento de F\'{\i}sica e Inform\'{a}tica, Instituto de F\'{\i}%
sica de S\~{a}o Carlos, Universidade de S\~{a}o Paulo, Caixa Postal 369,
13560-970 S\~{a}o Carlos, SP, Brazil\\
$^{2}$Centro Universit\'{a}rio da Funda\c{c}\~{a}o Educacional Guaxup\'{e} -UNIFEG,\\
Avenida Dona Floriana, 463, 37800-000 Guaxup\'{e}, MG, Brazil\\
$^{3}$Departamento de F\'{\i}sica, Universidade Federal de S\~{a}o Carlos, \\
Via Washington Luis Km
235, 13565-905, S\~{a}o Carlos,  SP, Brazil
}%

\date{\today}

\begin{abstract}
Atomic collisions are included in an interacting system of optical fields and trapped atoms
allowing field amplification. We study the effects of collisions on the system stability. Also a
study of the degree of entanglement between atomic and optical fields is made. We found that, for
an atomic field initially in a vacuum state and optical field in a coherent state, the degree of
entanglement does not depend on the optical field intensity or phase. We show that in conditions of
exponential instability the system presents at long times two distinct stationary degree of
entanglement with collisions affecting only one of them.

\end{abstract}

\maketitle

\section{introduction}\label{sec1}

Bose-Einstein condensation of trapped atomic gases \cite{r1a,r1b,r1c} has produced a fantastic
advance in atom optics. Particularly, the interaction of condensates with single mode quantized
light fields has been a fascinating topic
\cite{r2a,r2b,r2c,r2d,r2e,r2f,r2g,r2h,r10a,r10b,r10c,r11a,r11b,r12}, allowing for instance light
and matter wave amplification \cite{r10a,r10b,r10c}, optical control of atomic statistical
properties \cite{r11a,r11b}, and potential applications in quantum information technology
\cite{r12}.

It is known that the trap environment can modify the properties of ultra-cold atoms, such as its
critical temperature \cite{r13a,r13b}. Then it is expected that trap environment also influences
the interaction between ultra-cold atoms and optical fields. In Ref. \cite{r14}, the trap
environment effects on the Condensate Collective Atomic-Recoil Laser (CARL) \cite{r15a,r15b} was
considered by expanding the matter-wave field in the trap matter-wave modes. Such a situation was
called Cavity Atom Optics (CAO), in analogy with the Cavity Quantum Electrodynamics (CQE) where the
spontaneous emission is modified by the presence of the cavity. The model obtained presents
interesting properties such as two regimes of exponential instability\cite{r14} and statistical
properties that depend on intensity and phase of the optical field \cite{r16}. However, neither
atomic collisions was taken into account nor a detailed study of the degree of entanglement between
atomic and optical fields was done. Entanglement, one of the most notable characteristics of
quantum mechanics \cite{a1}, has been object of intense study, both in systems involving light and
matter \cite{a2a,a2b} and in solids \cite{a3a,a3b}, due to its important role for
quantum-information processing \cite{a4}. Therefore, inclusion of collisions and characterization
of entanglement are important in order to turn the model more realistic and potentially useful for
applications in the context of quantum information theory.

This paper deals with the extension of the model of atom-optical parametric amplifier considered in
Refs. \cite{r14,r16} by including collisions between atoms. A study of the changes due collisions
on the thresholds of exponential instability as well as in growth rate of the fields amplitudes is
presented. Also a characterization of the degree of entanglement between atomic and optical fields
is done. We show that in conditions of exponential instability the system presents at long times
two distinct stationary degree of entanglement with collisions affecting only one of them.

The article is organized as follows: In Section \ref{sec2} we derive the effective hamiltonian
describing the system studied. In Section \ref{sec3} we present an analysis of the system
stability. In Section \ref{sec4} we consider the atom-photon degree of entanglement in the regime
of field amplification. Finally, in Section \ref{sec5} we present the conclusion.

\section{model}\label{sec2}

We consider a Schr\"{o}dinger field of bosonic two-level atoms, with transition frequency $\nu$,
interacting via two-body collisions and coupled by electric-dipole interaction to two single-mode
running wave optical fields of frequencies $\omega _{1}$ and $\omega _{2}$, treated as a quantum
and a classical field, respectively. Both optical fields are assumed being far off-resonant from
any electronic transition. Thus, although the atom's internal state remains unchanged, the
center-of-mass motion may change due the atomic recoil induced by two-photon virtual transitions.
In the far off-resonance regime the excited state population is small, and therefore spontaneous
emission as well as collisions between excited state atoms, and between ground state and excited
states atoms, may be neglected. In this regime the excited state can be adiabatically eliminated
and the ground state atomic field evolves coherently under the effective Hamiltonian
\begin{widetext}
\begin{eqnarray} 
\hat{H}=\int d^{3}\mathbf{r\;}\hat{\Psi}^{\dagger }(\mathbf{r})\Bigg[
\mathcal{H}_{0}+\frac{U}{2}\hat{\Psi}^{\dagger }(\mathbf{r})\hat{\Psi}(%
\mathbf{r})+\hbar \left( \frac{g_{1}^{\ast }g_{2}}{\Delta }\hat{a}%
_{1}^{\dagger }a_{2}e^{-i\mathbf{K\cdot r+}}\frac{g_{2}^{\ast }g_{1}}{\Delta
}a_{2}^{\ast }\hat{a}_{1}e^{i\mathbf{K\cdot r}}\right) \Bigg] \hat{\Psi}(%
\mathbf{r})+ \hbar \delta\hat{a}_{1}^{\dagger }\hat{a}_{1}\label{eq1}
\end{eqnarray}
\end{widetext}
\normalsize
where
\begin{equation}
\mathcal{H}_{0}=-\frac{\hbar ^{2}}{2m}\nabla ^{2}+V(\mathbf{r}),  \label{eq2}
\end{equation}%
$m$ is the atomic mass, $V(\mathbf{r})$ is the trap potential, $\Delta = \omega_2 -\nu$ is the
detuning between atoms and the classical optical field, $g_{1}$ and $g_{2}$
are the optical fields coupling coefficients, $\mathbf{K}=%
\mathbf{k}_{1}-\mathbf{k}_{2}$ is the difference between their wavevectors, and $\delta =\omega
_{1}-\omega _{2}$ is the detuning between them. The operator $\hat{a}_{1}$ is the photon
annihilation operator
for the quantized optical field, taken in the frame rotating at the classical field frequency $%
\omega _{2}$. The optical field treated classically is assumed to remain undepleted, so that
$a_{2}$ is simply a constant related to its intensity. Terms corresponding to the spatially
independent light shift potential were neglected, so that the index of refraction of the atomic
sample was assumed the same of the vacuum. Two-body collisions were included in the s-wave
scattering limit by the second term inside brackets in the  Hamiltonian (\ref{eq1}), where
\begin{equation}
U=\frac{4\pi \hbar ^{2}a}{m},  \label{eq3}
\end{equation}%
and $a$ is the s-wave scattering length, which depending on the repulsive or attractive character
of the interaction can assume positive or negative values, respectively. We consider only positive
values of scattering length, which are suitable for the creation of large condensates, and
correspond to a situation more consistent with the approximations in this paper.

We assume that the atomic field is initially a Bose-Einstein condensate with mean number of
condensed atoms $N$, and that this condensate is well described by a number state so its initial
state is described by
\begin{equation}
\left\vert \psi (t=0)\right\rangle =\frac{1}{\sqrt{N!}}(\hat{c}_{0}^{\dagger })^{N}\;\left\vert
0\right\rangle  \label{eq4}
\end{equation}%
where $\left\vert 0\right\rangle $ is the vacuum state, and
\begin{equation}
\hat{c}_{0}^{\dagger }=\int d^{3}\mathbf{r\;}\varphi _{0}(\mathbf{r)}\hat{%
\Psi}^{\dagger }(\mathbf{r})  \label{eq5}
\end{equation}%
is the creation operator for atoms in the condensate state $\varphi _{0}(%
\mathbf{r)}$. Due to the presence of collisions the condensate wave function $%
\varphi _{0}(\mathbf{r)}$ satisfies the Gross-Pitaevskii equation \cite{gpa,gpb}
\begin{equation}
\left( \mathcal{H}_{0}+N U\left\vert \varphi _{0}(\mathbf{r)}%
\right\vert ^{2}\right) \varphi _{0}(\mathbf{r)=}\mu \varphi _{0}(\mathbf{r),%
}  \label{eq6}
\end{equation}%
where $\mu $ is the chemical potential.

Now we expand the atomic field operator in terms of the trap eigenmodes $%
\left\{ \varphi _{n}(\mathbf{r})\right\} $ according to%
\begin{equation}
\hat{\Psi}(\mathbf{r})=\varphi _{0}(\mathbf{r})\,\hat{c}_{0}+\delta \hat{\psi%
}(\mathbf{r}),  \label{eq7}
\end{equation}%
where $\delta \hat{\psi}(\mathbf{r})=\sum_{n\neq 0}^{\infty }\varphi _{n}(%
\mathbf{r})\,\hat{c}_{n}$ creates particles in the excited trap modes, $\hat{c}_{n}$ is the
annihilation operator for atoms in mode $n$, and $\int d^{3}\mathbf{r\;}\varphi _{m}^{\ast }(\mathbf{r)%
}\varphi _{n}(\mathbf{r)}=\delta _{mn}.$
We are interested in the linear regime, valid for interaction times so that $\left\langle \int d^{3}\mathbf{%
r\;}\delta \hat{\psi}^{\dagger }\delta \hat{\psi}\right\rangle \ll \left\langle
\hat{c}_{0}^{\dagger }\hat{c}_{0}\right\rangle $. Then we can maintain only quadratic terms in the
fields operators and invoke \ the undepleted approximation, which permits to substitute the
condensate mode by a c-number evolving as
\begin{equation}
c_{0}(t)\approx \sqrt{N}e^{-i\mathbf{\mu }t}.  \label{eq11}
\end{equation}%
Therefore, in the linear regime, by inserting expansion (\ref{eq7}) and taking into account the
Gross-Pitaevskii equation (Eq. (\ref{eq6})), the Hamiltonian (\ref{eq1}) reduces to
\begin{widetext}
\begin{eqnarray}
\hat{H} &=&\hbar \delta \hat{a}_{1}^{\dagger }\hat{a}_{1}+\hbar \int d^{3}%
\mathbf{r\;}\delta \hat{\psi}^{\dagger }\left( \mathcal{H}_{0}+2NU\right)
\delta \hat{\psi}+\hbar \frac{U}{2}\left( Ne^{2i\mathbf{\mu }t}\int d^{3}%
\mathbf{r\;}\varphi _{0}^{*2}\delta \hat{\psi}^{2}+h.c.\right)\nonumber\\
&+&\hbar \left[\frac{g_{1}^{\ast }g_{2}a_{2}}{\Delta }\sqrt{N}\hat{a}_{1}^{\dagger }\mathbf{\int
}d^{3}\mathbf{r}\left(e^{i\mathbf{\mu }t}\varphi _{0}^{\ast }e^{-i%
\mathbf{K\cdot r}}\delta \hat{\psi}+e^{-i\mathbf{\mu }t}\delta\hat{\psi}^{\dagger }e^{-i%
\mathbf{K\cdot r}}\varphi _{0}  \right)+h.c.\right], \newline \label{eq12}
\end{eqnarray}%
\end{widetext}
which in terms of the trap excited modes expansion becomes
\begin{eqnarray}
\hat{H}&=&\hbar \delta \hat{a}^{\dagger }\hat{a}+\hbar \sum_{n\neq 0}\omega _{n}\hat{c}%
_{n}^{\dagger }\hat{c}_{n}
+\hbar \sum_{nl\neq 0}\frac{\kappa _{nl}}{2}(\hat{c}_{n}\hat{c}_{l}
+\hat{c}_{n}^{\dagger}\hat{c%
}_{l}^{\dagger})\nonumber\\
&&+\hbar\left( \hat{a}+\hat{a}^{\dagger} \right)\sum_{n\neq 0}\chi
_{n}\left(\hat{c}_{n}+\hat{c}_{n}^{\dagger }\right) , \label{lrt}
\end{eqnarray}
where
\begin{equation}
\hbar \omega _{n}=\int d^{3}\mathbf{r}\varphi _{n}^{\ast }\left( \mathcal{H}_{0}+2NU\left\vert
\varphi _{0}(\mathbf{r)}\right\vert ^{2}\right) \varphi _{n} ,
\end{equation}%
is the collision modified energy of the $n$th trap level, $\chi _{n}=\sqrt{N}%
A_{0n}\left\vert g_{1}\right\vert \left\vert g_{2}\right\vert \left\vert a_{2}\right\vert
/\left\vert \Delta \right\vert $ is the effective coupling
constant between the condensate and the quantum optical field, $\hat{a}%
=(g_{1}g_{2}^{\ast }a_{2}^{\ast }\Delta /\left\vert g_{1}\right\vert \left\vert g_{2}\right\vert
\left\vert a_{2}\right\vert \left\vert \Delta \right\vert )\hat{a}_{1}$ is the optical field
operator multiplied by a phase factor that is related to the classical optical field phase,

\begin{equation}
\kappa _{nl}=\frac{2NU}{\hbar }\mathbf{\int }d^{3}\mathbf{r}\varphi _{0}^{2}\varphi _{n}^{\ast
}\varphi _{l}^{\ast } \label{overlap}
\end{equation}%
and
\begin{equation}
A_{0 n}=\mathbf{\int }d^{3}\mathbf{r}\varphi _{n}^{\ast }e^{-i\mathbf{%
K\cdot r}}\varphi _{0} \label{optical}
\end{equation}%
are the collision parameter and the element of matrix for the optical transition, respectively,
both assumed being real numbers. The phase factors $e^{\pm i\mathbf{\mu }t}$ were included in the
operators $\hat{c}_{n}$ and $\hat{c}_{n}^{\dagger }$.

For simplicity, we assume that the overlap-integral in the collision parameter given by Eq.
(\ref{overlap}) has a significant contribution for a given $n=l=m$ and the matrix for optical
transition $A_{0 n}$ is sharply peaked for $n=m$ \footnote{This could be in principle obtained in
Fabry-P{\'e}rot-type matter-wave resonators \cite{wilkens}, where the absolute value of momentum is
relatively well defined for each trap level.}. In this case we can neglect all excited trap modes
except the $m$-mode in the Hamiltonian (\ref{lrt}), and the following effective Hamiltonian is
obtained
\begin{eqnarray}
\hat{H}&=&\hbar \delta \hat{a}^{\dagger }\hat{a}+\hbar \omega _{m}\hat{c}%
_{m}^{\dagger }\hat{c}_{m}+\hbar \frac{\kappa _{m}}{2}(\hat{c}_{m}^{2}
+\hat{c%
}_{m}^{\dagger 2})\nonumber\\
&&+\hbar \chi _{m}\left( \hat{a}^{\dagger }\hat{c}%
_{m}^{\dagger }+\hat{a}^{\dagger }\hat{c}_{m}+\hat{c}_{m}^{\dagger }\hat{a}+%
\hat{c}_{m}\hat{a}\right) ,  \label{lr3}
\end{eqnarray}

Terms like $\hat{a}^{\dagger }\hat{c}_{m}^{\dagger }$ in Hamiltonian (\ref{lr3}) correspond to the
generation of correlated atom-photon pair. These terms are analogous to the Nondegenerated Optical
Parametric Amplifier \cite{walls}. Inclusion of collisions introduces terms like $\hat{c}_{m}^{\dag
2}$, which are responsible by creation of an atomic pair, and are analogous to the Degenerated
Optical Parametric Amplifier \cite{walls}. The Heisenberg equations of motion for
the field operators obtained from Hamiltonian (\ref{lr3}) result the following $%
4\times 4$ linear system of equations
\begin{equation}
\frac{d}{dt}\left(
\begin{array}{c}
\hat{c} \\
\hat{c}^{\dagger } \\
\hat{a} \\
\hat{a}^{\dagger }%
\end{array}%
\right) =i\left(
\begin{array}{cccc}
-1 & -\kappa & -\chi & -\chi \\
\kappa & 1 & \chi & \chi \\
-\chi & -\chi & -\delta & 0 \\
\chi & \chi & 0 & \delta%
\end{array}%
\right) \left(
\begin{array}{c}
\hat{c} \\
\hat{c}^{\dagger } \\
\hat{a} \\
\hat{a}^{\dagger }%
\end{array}%
\right) ,  \label{lr2}
\end{equation}%
where the index $m$ was dropped in order to simplify notation, and we introduced the dimensionless
parameters $t=\omega _{m}t$, $\delta =\delta /\omega _{m}$, $\kappa =\kappa _{m}/\omega _{m}$, and
$\chi =\chi _{m}/\omega_{m} $.

\section{stability}\label{sec3}
The solution of the linear system (\ref{lr2}) can be written as
\begin{equation}
\hat{x}_{i}(t)=\sum_{j=1}^{4}G_{ij}(t)\,\hat{x}_{j}(0),  \label{sum}
\end{equation}%
where we defined $\hat{x}_{1}=\hat{c}$, $\hat{x}_{2}=\hat{c}^{\dagger }$, $\hat{x}_{3}=\hat{a}$ and
$\hat{x}_{4}=a^{\dagger }$ for
convenience, $G_{ij}(t)=\sum_{k=1}^{4}\left[ \mathbf{U}\right] _{ik}[\mathbf{%
U}^{-1}]_{kj}e^{i\mathbf{\omega }_{k}t}$, $\left[ \mathbf{U}\right] _{ik}$ is \ the $i$th component
of the $k$th eigenvector of the matrix at RHS of Eq. (\ref{lr2}), and $\omega _{k}$ are the system
eigenfrequencies.

Stability analysis shows two regimes of exponential instability: (i) For $%
\delta >0$ and $\chi ^{2}>\delta (1+\kappa )/4$ there are two purely real and two purely imaginary
eigenfrequencies of the form $\left\{ \omega _{1}=\Omega ,\omega _{2}=-\Omega ,\omega _{3}=i\Gamma
,\omega _{4}=-i\Gamma \right\} $, where $\Omega $ and $\Gamma $ are both real quantities. There is
only one exponentially growing solution at the imaginary frequency $\omega _{4}$ and the system is
unstable. (ii) For $\delta <0$ and $\chi ^{2}>(1-\kappa ^{2}-\delta ^{2})^{2}/16\left\vert \delta
\right\vert \left(
1-\kappa \right) $ the eigenfrequencies are complex numbers of the form \small $%
\left\{ \omega _{1}=\Omega +i\Gamma,\hspace{0.1cm}\omega _{2}=-\Omega +i\Gamma,\hspace{0.1cm}\omega _{3}=\Omega -i\Gamma,\hspace{0.1cm}\omega _{4}=-\Omega -i\Gamma \right\} $. \normalsize This case
presents two exponentially growing solutions, $\omega _{3}$ and $\omega _{4}$%
, which grow at the same rate $\Gamma $, but rotate at equal and opposite frequencies $\pm \Omega
$, producing a beating in the exponential growth of the fields intensities. Otherwise the
eigenfrequencies are real and the system is stable.

\begin{figure}[h!]
\includegraphics[height=6cm]{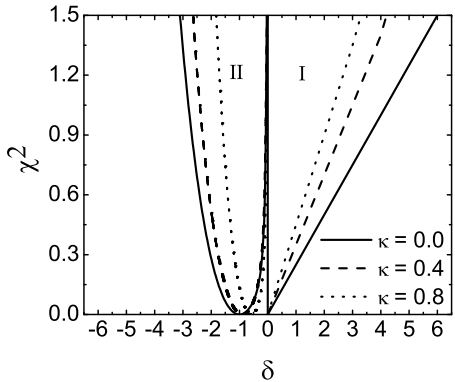}
\caption{\label{f1} Instability domains. Points inside region I correspond to one exponentially
growing solution, whereas for points inside region II there are two counter-rotating exponentially
growing solutions. The full line define the threshold in the absence of collisions ($\kappa =0.0$).
Change in the threshold due collisions correspond to the dashed ($\kappa=0.4$) and dotted lines
($\kappa = 0.8$). The parameters are dimensionless.}
\end{figure}

In Fig. \ref{f1} we plot in the $\delta -\chi^{2}$ plane the values of parameters defining the
threshold between stable and unstable solutions. Points inside region I correspond to one
exponentially growing solution, whereas for points inside region II there are two counter-rotating
exponentially growing solutions. The full line indicates the threshold in the absence of collisions
whereas dashed and dotted lines shows the change due collisions, which have the effect of reducing
the unstable regions. In addition, for small $\chi^2$ the region II is centered around $\delta
=-\sqrt{1-\kappa^2}$ and at asymptotically large detunings its threshold grows as $\chi ^{2}>
-\delta^{3}/16\left( 1-\kappa \right)$.

\begin{figure}
\includegraphics[height=10cm]{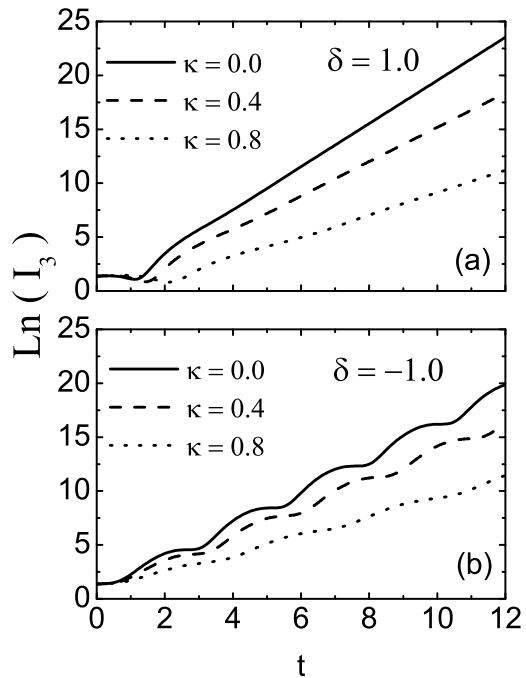}
\caption{\label{f2} Logarithmic plot of the light field intensity as function of time. (a) for
parameters lying over the region I of the instability domains. (b) for parameters lying over the
region II of the instability domains. Full line corresponds to absence of collisions ($\kappa
=0.0$) while dashed and doted lines corresponds to inclusion of collisions with $\kappa =0.4$ and
$\kappa =0.8$, respectively.  We set $\chi =1$ and $\alpha =2$. The parameters are dimensionless. }
\end{figure}

In order to illustrate the effects of collisions on the fields intensities $%
I_{i}(t)=\left\langle \hat{x}_{i}^{\dagger }(t)\hat{x}_{i}(t)\right\rangle$, with $i=1,3$, we
consider that the atomic mode begins in a vacuum state $|0\rangle$ whereas the light field is
initially in a coherent state $\left\vert \alpha \right\rangle $. Then, with the help of Eq.
(\ref{sum}), we found
\begin{equation}
I_{i}(t)=\left\vert G_{i2}(t)\right\vert ^{2}+\left\vert G_{i4}(t)\right\vert ^{2}+\left\vert
G_{i3}(t)\alpha +G_{i4}(t)\alpha ^{\ast }\right\vert ^{2}.  \label{inten}
\end{equation}%
Figs. \ref{f2}-(a) and \ref{f2}-(b) shows the logarithmic plot of the optical field intensity
$I_{3}(t)$ as a function of time for parameters lying in the regions I and II of instability,
respectively. The intensity of the atomic mode has a similar behavior. We see that collisions
reduce the rate of exponential growing and change the beating frequency of oscillation of the field
intensity. The change in the density distribution of atoms in the condensate mode due to collisions
reduces the scattering of photons into the optical field, as well as the scattering of atoms by the
optical field into the atomic mode, in opposition to field amplification.

\section{entanglement}\label{sec4}

Now we turn to analyze the atom-photon degree of entanglement for the exponential growing regimes
of fields amplitudes. It is known that for a two-component system in a pure state the degree of
entanglement can be characterized by the entropy or purity of one of the system components
\cite{pega,pegb}. Such entanglement measures require the calculation of the time-dependent quantum
state for the system, which may not be an easy task. Instead, we consider a recently proposed
entanglement coefficient in terms of cross-covariances of the fields operators defined by
\cite{dodonov}

\begin{equation}
Y=\left[ \frac{\left\vert \overline{\hat{a}\hat{c}^{\dagger }}\right\vert
^{2}+\left\vert \overline{\hat{a}\hat{c}}\right\vert ^{2}}{2\left( \overline{%
\hat{a}^{\dagger }\hat{a}}+\frac{1}{2}\right) \left( \overline{\hat{c}%
^{\dagger }\hat{c}}+\frac{1}{2}\right) }\right] ^{1/2},  \label{335}
\end{equation}%
with the notation $\overline{x_{i}x_{j}}=\left\langle x_{i}x_{j}\right\rangle -\left\langle
x_{i}\right\rangle \left\langle x_{j}\right\rangle $ for the unsymetrized centered second-order
moments of the operators. Since we know the time dependent solutions for the fields amplitudes the
parameter $Y$ is easily calculated for the system considered in this paper.

\begin{figure}
\includegraphics[height=7cm]{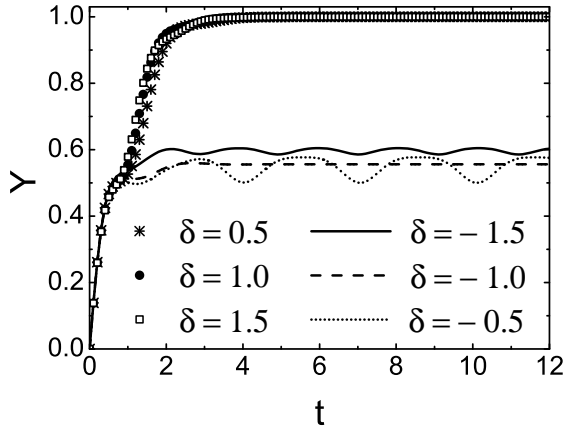}
\caption{\label{f3}Plot of the entanglement coefficient as a function of time in absence of
collisions for several values of detuning $\delta$. We set $\chi =1$. The parameters are
dimensionless.}
\end{figure}

The coefficient $Y$ satisfies the inequality $0\leq Y<1$, where at the maximum value of $Y$ the
system is maximally entangled and was introduced taking into account that for two systems with
operators $\hat{A}$ and $\hat{B}$, if any of the cross-covariances with these operators differs
from zero, then the modes are entangled. For the excited trap mode starting in a vacuum state and
the light field in a coherent state, we obtain,
\begin{widetext}
\begin{equation}
Y=\left[ \frac{\left\vert G_{31}(t)G_{11}^{\ast }(t)+G_{33}(t)G_{13}^{\ast }(t)\right\vert
^{2}+\left\vert G_{31}(t)G_{12}(t)+G_{33}(t)G_{14}(t)\right\vert ^{2}}{2\left( \left\vert
G_{32}(t)\right\vert ^{2}+\left\vert G_{34}(t)\right\vert ^{2}+\frac{1}{2}%
\right) \left( \left\vert G_{12}(t)\right\vert ^{2}+\left\vert G_{14}(t)\right\vert
^{2}+\frac{1}{2}\right) }\right] ^{1/2}.  \label{237*}
\end{equation}%
\end{widetext}
Eq. (\ref{237*}) shows that the degree of entanglement do not depend on the optical field intensity
or phase. Furthermore, in the regime of exponential instability the parameter $Y$ attains at long
times a stationary value thats is dependent on the signal of the detuning $\delta $. To see that,
in Fig. \ref {f3} the parameter $Y$ is plotted as a function of time for several values of
detunings in the absence of collisions. We observe that for $\delta >0$ (one exponentially growing
solution) the system attains at long times the maximum degree of entanglement, while for $\delta
<0$ (two exponentially growing counter-rotating solutions) the long time degree of entanglement
attains an oscillating stationary value far below the maximum one.

\begin{figure}
\includegraphics[height=10cm]{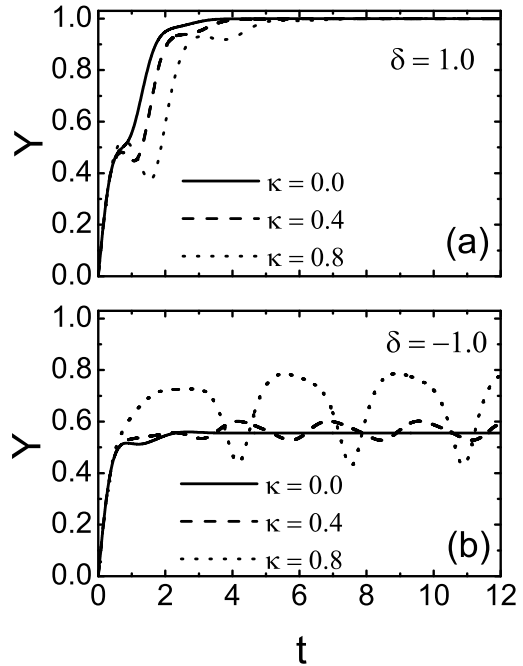}
\caption{\label{f4}Plot of the entanglement coefficient as a function of time. (a) for parameters
lying over the region I of the instability domains. (b) for parameters lying over the region II of
the instability domains. Full line corresponds to absence of collisions ($\kappa =0.0$) while
dashed and doted lines corresponds to inclusion of collisions with $\kappa =0.4$ and $\kappa =0.8$,
respectively. We set $\chi =1$. The parameters are dimensionless.}
\end{figure}

The effects of collisions are presented in Figures \ref{f4}-(a) and \ref {f4}-(b) that show the
entanglement coefficient as a function of time for values of parameters lying in regions of one and
two exponential growing solutions, respectively, and considering different values for the collision
parameter $\kappa $. We see in Fig. \ref{f4}-(a) that collisions do not affect at long times the
degree of entanglement for $\delta>0,$ which always tend to the maximum value. However, for $\delta
<0$ (Fig. \ref{f4}-(b)) collisions strongly affects the long time degree of entanglement by
changing the amplitude of oscillations of its stationary value.

\section{conclusion}\label{sec5}

In conclusion, in this work we have included atomic collisions in a model of an atom-optical
parametric amplifier of trapped atoms. Analyzing the system stability in the regime of field
amplification, we found that atomic collisions reduce the growth rate of the fields amplitudes. For
an atomic field initially in a vacuum state and optical field in a coherent state, we have verified
that the degree of entanglement between atomic and optical fields does not depend on the optical
field intensity or phase. Furthermore, in conditions of field amplification the degree of
entanglement attains at long time a stationary value dependent on the regime of exponential
instability, being maximum only in the case of one exponentially growing solution. Finally, atomic
collisions only affect the long time degree of entanglement in the case of two exponentially
growing solutions.

\begin{acknowledgments}
This work was supported by CAPES (DF, Brazil) and FAPESP (SP, Brazil). We are grateful to Klaus
Capelle for useful suggestions.
\end{acknowledgments}


\begin{thebibliography}{99}

\bibitem{r1a} M. H. Anderson \textit{et al}, Science \textbf{269}, 198
(1995).

\bibitem{r1b} K. B. Davis \textit{et al}, Phys. Rev. Lett. \textbf{75}, 3969 (1995).

\bibitem{r1c} C. C. Bradley, C. A. Sackett, J. J. Tollett, and R. G. Hulet, Phys. Rev. Lett. \textbf{75}, 1687 (1995).

\bibitem{r2a} K-P Marzlin and J. Audretsch, Phys. Rev. A \textbf{57}, 1333 (1998).

\bibitem{r2b} H. J. Wang, X. X. Yi, and X. W. Ba, Phys. Rev. A \textbf{62}, 023601 (2000).

\bibitem{r2c} S. Inouye \textit{et al}, Phys. Rev. Lett. \textbf{85}, 4225 (2000).

\bibitem{r2d} P. Horak and H. Ritsch, Phys. Rev. A \textbf{63}, 023603 (2001).

\bibitem{r2e} D. Jaksch, S. A. Gardiner, K. Schulze, J. I.Cirac, and P. Zoller, Phys. Rev. Lett.\textbf{86}, 4733 (2001).

\bibitem{r2f} M. K. Olsen, J. J. Hope, and L. I. Plimak, Phys. Rev. A \textbf{64}, 013601 (2001).

\bibitem{r2g} C. P. Search and P. R. Berman, Phys. Rev. A \textbf{64,} 043602 (2001).

\bibitem{r2h} C. P. Search, Phys. Rev. A \textbf{64,} 053606 (2001).

\bibitem{r10a} C. K. Law and N. P. Bigelow, Phys. Rev. A \textbf{58}, 4791
(1998).

\bibitem{r10b} M. G. Moore, O. Zobay, and P. Meystre, Phys. Rev. A \textbf{60}%
, 1491 (1999).

\bibitem{r10c} G. A. Prataviera and E. M. S. Ribeiro, Phys. Rev. A \textbf{65}%
, 033622 (2002).

\bibitem{r11a} H. Zeng, F. Lin, and W. Zhang, Phys. Lett. A \textbf{201}, 397
(1995).

\bibitem{r11b} M. G. Moore and P. Meystre, Phys. Rev. A \textbf{59}, R1754 (1999).

\bibitem{r12} G. A. Prataviera and M. C. de Oliveira, Phys. Rev. A \textbf{70}%
, 011602(R) (2004).

\bibitem{r13a} V. Bagnato, D. E. Pritchard, and D. Kleppner, Phys. Rev. A \textbf{35}, 4354 (1987).

\bibitem{r13b} C. P. Search, H. Pu, W. Zhang, B. P. Anderson, and P. Meystre, Phys. Rev. A \textbf{65},
063616 (2002).

\bibitem{r14} J. Heurich, M. G. Moore and P. Meystre, Opt. Comm. \textbf{179}%
, 549 (2000).

\bibitem{r15a} R. Bonifacio, L. De Salvo, L. M. Narducci, and E. J. DAngelo, Phys. Rev. A
\textbf{50}, 1716 (1994).

\bibitem{r15b} M. G. Moore and P. Meystre, \textit{ibid.} \textbf{58}, 3248 (1998).

\bibitem{r16} G. A. Prataviera, Phys. Rev. A \textbf{67}, 045602 (2003).

\bibitem{a1}A. Einstein, B. Podolsky, and N. Rosen, Phys. Rev. \textbf{47}, 777 (1935).

\bibitem{a2a} H. Haffner, W. Hansel, C. F. Roos; et al., Nature \textbf{438}, 643 (2005).

\bibitem{a2b} R. N. Palmer, C. MouraAlves, D. Jaksch, Phys. Rev. A \textbf{72}, 042335 (2005).

\bibitem{a3a} P. Zanardi, Phys. Rev. A \textbf{65}, 042101 (2002)

\bibitem{a3b} V. V. Fran\c{c}a, K. Capelle, Phys. Rev. A \textbf{74}, 042325 (2006).

\bibitem{a4} C. H. Bennett, D. P. Di Vincenzo, Nature 404, 247 (2000).

\bibitem{gpa} E. P. Gross, Nuevo Cimento \textbf{20}, 454 (1961).

\bibitem{gpb} L. P. Pitaevskii, Zh. {\'E}ksp. Teor. Fiz. \textbf{40},646 (1961) [Sov. Phys. JETP \textbf{13}, 451
(1961)].

\bibitem{walls} D. F. Walls and G. J. Milburn, \textit{Quantum Optics} (Springer-Verlag, Berlin, 1994).

\bibitem{wilkens} M. Wilkens,E. Goldstein, B. Taylor, and P. Meystre ,Phys. Rev. A \textbf{47}, 2366 (1993).

\bibitem{pega} S. M. Barnett, and S. J. D. Phoenix, Phys. Rev. A \textbf{40},2404 (1989).

\bibitem{pegb} W. H. Zurek, S. Habib, and J. P. Paz, Phys. Rev. Lett. \textbf{70}, 1187(1993).

\bibitem{dodonov} V. V. Dodonov, A. S. M. de Castro, and S. S. Mizrahi, Phys. Lett. A,
\textbf{296}, 73 (2002).


\end{thebibliography}

\end{document}